# Measuring thermal conductivity in extreme conditions: sub-Kelvin temperatures and high (27 T) magnetic fields

P. J. E. M. van der Linden[*]
Grenoble High Magnetic Field Laboratory (CNRS-MPI), BP 166, F- 38042 Grenoble, France

K. Behnia
Laboratoire de Physique Quantique (CNRS), ESPCI, 10 Rue Vauquelin, F-75005 Paris, France

**Abstract**

We present a one-heater-two-thermometer set-up for measuring thermal conductivity and electric resistivity of a bulk sample at low temperatures down to 0.1 K and in magnetic fields up to 27 Tesla. The design overcomes the difficulties emerging in the context of large water-cooled resistive magnets.

**Introduction**

Thermal conductivity has proved to be a useful tool for investigating the electronic properties of metals and superconductors [1-6]. For example, in superconductors, the study of heat transport in the superconducting state can give information on electron scattering mechanisms. Obviously, this information is not accessible by measuring the electric resistivity. However, heat is carried both by electrons and phonons and separating the two contributions is seldom straightforward. In many cases, it is desirable to cool the sample down to very low temperatures to reach the well-understood asymptotic behavior of heat conductivity for both electronic and phononic components. Supplementary information can be obtained by the application of a strong magnetic field. In superconductors, one can destroy superconductivity and compare thermal conductivity in the normal and superconducting states [3-5]. Measuring thermal conductivity in these extreme conditions is not a trivial task. Here, we describe a set-up used to measure the thermal conductivity of a solid in temperatures down to 0.1K in presence of a magnetic field of 27T produced by a water-cooled resistive magnet.

In the standard set-up for measuring thermal conductivity working in the sub-Kelvin range, the sample is held in vacuum and thermally attached to the cold finger of a bottom-loading dilution refrigerator. Miniature (< 1mm$^2$) resistor chips are attached to electrodes on the sample in order to inject heat and to measure local temperature profile across the sample. This set-up has been successfully used in magnetic fields produced by a superconducting magnet [1-5]. However, in order to use such a configuration to measure sub-Kelvin thermal conductivity in magnetic fields larger than 20T, one should overcome two difficulties.

First of all, a magnetic field of such a magnitude is currently obtained only with resistive magnets cooled by a large flow of water [7]. Strong mechanical vibrations associated with water circulation constitute an unwelcome source of heat and a tremendous obstacle for cooling the sample and the thermometers to very low temperatures. In a first attempt, we used a bottom-loading dilution refrigerator and found that because of these vibrations, the sample, with its weak thermal link to the dilute phase inside the mixing chamber, could not be cooled down below 300 mK in high magnetic fields. To circumvent this limitation we have developed a set-up in which a vacuum chamber containing the sample and thermometers can be introduced into the mixing chamber of a top-loading dilution refrigerator.

The second challenge is to accurately measure the temperature in this temperature and field range, where the magneto-resistance of thermometers is far from negligible. In the case of superconducting magnets, it is possible to create a zero-field zone at a short distance from the maximum-field region and to use this "compensated zone" for calibration. This option does not exist for the large resistive magnets. For this purpose, we used an alternative recent technology called Coulomb Blockade Thermometry [8]. A Coulomb Blockade Thermometer (CBT) provided by Nanoway (Finland), a primary thermometer known to show no detectable variation with magnetic field [9], was used for an *in situ* calibration of our resistive thermometers.

This set-up was used to measure the thermal conductivity of a high-$T_c$ single crystal at 25T for the first time [10].

**Experimental setup**

Figure 1 shows a schematic diagram of the experimental setup to give an idea of the dimensions of the experimental environment. The 20 MW resistive magnet capable of producing 30 Tesla is described elsewhere [7]. The dilution refrigerator is introduced into the 50 mm room temperature magnet bore. The cryostat lift combined with the XY table allow for linear movements over three axes and tilt over two axes. Thus, the 47mm cryostat tail can be aligned in the magnet in such a way that the two do not touch each other. This is essential to reduce the mechanical vibrations associated with water circulation.

---

[*] Present address: ESRF, BP 220, F-38043 Grenoble cedex, France

The refrigerator is a commercial Oxford Instruments Kelvinox unit inserted into a purpose-built $^4$He bath cryostat. This top-loading dilution refrigerator has a base temperature of 20 mK and cooling power of 200 µW at 100 mK at the sample position. It has a 6 mm central access hole from room temperature into the mixing chamber to be used for services such as a rotating axis, optical fibers, a high frequency coaxial [11] or, in this case, a vacuum pumping tube.

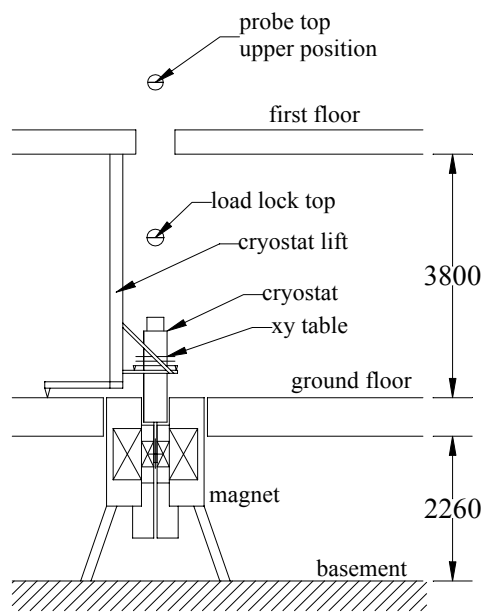

Fig.1- A schematic view of the experiment

The mixing chamber diameter available for the experiment is 24 mm. A detailed exploded view of the vacuum chamber is represented in figure 2. The different parts are machined in Araldite epoxy and assembled using Stycast 1266. The vacuum chamber cork on which the experiment is mounted is used as a feed through for the wiring and the 125 µm silver foil cold finger. The cold finger has been designed in order to reduce eddy current heating due to the unavoidable field ripple. To improve heat exchange between the dilute liquid and the silver the bottom part of the cold finger was coated with a 1.5mm layer of sintered silver. The fabrication process does not allow for much thinner layers. The eddy current heating in the silver was estimated at a few microWatts. The vacuum chamber isolation vacuum is pumped at room temperature. Since the chamber is immersed in the dilute phase of the mixture, the introduction of Helium exchange gas during cool down is unnecessary and therefore the isolation vacuum is of very high quality.

Inside the vacuum chamber the cold finger is equipped with a Ruthenium Oxide thermometer, a CBT [6] and a metal-film resistance heater. In this way, the sample temperature can be varied from base temperature up to 1K while the dilution refrigerator runs at or slightly above base temperature.

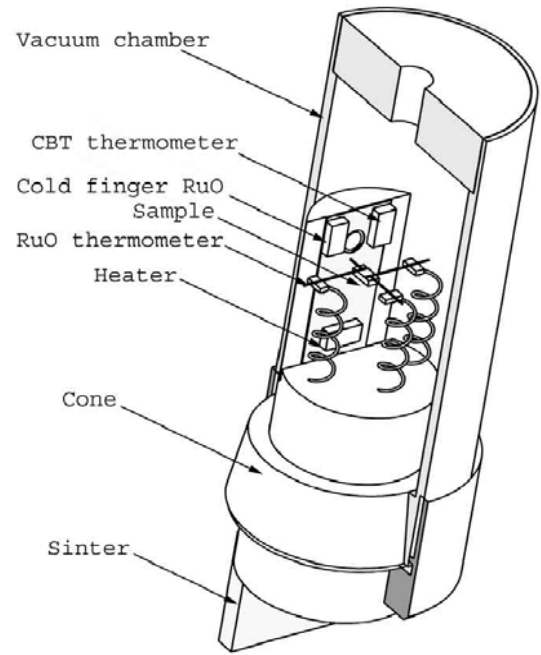

Fig.2- The vacuum chamber and the thermal conductivity set-up

The cold finger supports a standard four-probe heat conductivity measurement set-up. It consists of three Ruthenium Oxide resistors, which are thermally attached to three electrodes evaporated on the sample. A fourth electrode connects one end of the sample to the cold finger. These links have been made using 25µm gold wire and Dupont 4029 silver epoxy. Note that these thermal links through metallic wires can be used to measure the electric resistivity of the sample. Hence, during one cooling run, both the thermal conductivity and the electrical resistivity may be measured. These gold wires are a fragile part of the set-up; if they are too short the thermal stresses during cool down will break the silver paint contact and if they are too long the setup becomes very sensitive to mechanical vibrations. Special care was paid to reduce the thermal exchange between the set-up and the outside world. In order to inject current into the sample, the two thermometers and the heater and also to measure the produced voltages, we used 50µm diameter manganin wire. The wires were pair-twisted in order to reduce the noise induced by field ripples. They were then coiled up to be able to support the weight of RuO$_2$ chips. In order to reduce the sensitivity of these coils to mechanical vibrations, the three resistor chips were anchored with thin (2μm) Kevlar fibers.

Six twisted pairs of 75 μm manganin wire and six twisted pairs of 62μm Phosphorus Bronze wire were used to connect the set-up to a series of electrical relays on a thermal bath situated on the still. From this point to the room temperatures, twisted pairs of 100 μm copper wires were used. In order to avoid heating of the thermometers by ambient Radio Frequency (RF), signals were RF filtered at the room temperature connector using Murata NFE61P 4700 pF filters.

A steady-state method [12, 13] was employed for our measurements. For each measurement, the magnetic field as well as the temperature of the cold finger was kept constant. Then, a finite current was applied to the heater leading to the gradual emergence of a temperature gradient across the sample. The time needed to attain the steady state was found to increase from a fraction of minute to several minutes at the lowest temperatures. In the steady state, the resistance of the two thermometers was measured with a couple of homebuilt fixed-gain (A=100) low frequency differential preamplifiers based on Burr-Brown OPA27 and INA105 integrated circuits. This allowed us to avoid noise pickup over the six meters long coaxial cables. Moreover, by measuring the resistance difference of the two thermometers, one could directly have access to the temperature difference. Standard digital lock-in amplifiers (Standford SR830) were used for measuring the resistance of the sample and/or the thermometers. A Keithley 2400 source-meter was used to power the heater. The same instruments were used to measure the differential conductivity of the CBT at fixed bias voltage, which allowed us to calibrate the two thermometers in a magnetic field.

## Results

In order to check the accuracy of the thermometry in presence of a strong magnetic field, the set-up was used to measure the low-temperature thermal conductivity of a gold wire (purity: 99.9 %, diameter: 17 μm, length: 10.5 mm). Fig. 3 displays the temperature dependence of thermal conductivity, κ, for zero magnetic field and for H= 25T. As seen in the figure, in both cases, thermal conductivity divided by temperature, κ/T, is constant in the explored temperature window. In other words, thermal conductivity is a purely linear function of temperature. The application of a magnetic field leads to a threefold decrease in the magnitude of this linear thermal conductivity.

The conduction of heat in a metal such as gold is totally dominated by the contribution of itinerant electrons. A linear thermal conductivity is thus naturally expected at low temperatures, when the electronic mean-free-path becomes temperature-independent. The magnitude of this linear term can be compared to what is expected according to the Wiedemann-Franz law which strictly relates the conduction of heat and charge by electrons in a metal.

According to this law, which prevails when electrons are only scattered by defects and impurities of the solid, the ratio of thermal (κ) to electrical conductivity (σ) divided by temperature is equal to a universal constant:

$$\kappa/\sigma T = \kappa \rho / T = L_0$$

$L_0$, also known as the Sommerfeld number, is related to the charge of electron, e, and the Boltzmann constant $k_B$:

$$L_0 = \pi^2/3 \, (k_B/e)^2 = 24.4 \times 10^{-9} \, V^2/K^2$$

As seen in the Figure, both at zero magnetic filed and at H=25T, the magnitude of κ/T is very close to $L_0/\rho_0$ where $\rho_0$ is the measured residual resistivity of the wire. Thus, the threefold decrease in κ/T faithfully reflects the positive magneto-resistance of gold at these temperatures. The small mismatch between the magnitudes of κ/T and $L_0/\rho_0$ (1 percent at zero-field and 3 percent at H=25T) allows us to quantify our precision in measuring the absolute magnitude of κ/T. This confirms the robustness of Coulomb Blockade thermometry in presence of a strong magnetic field [14].

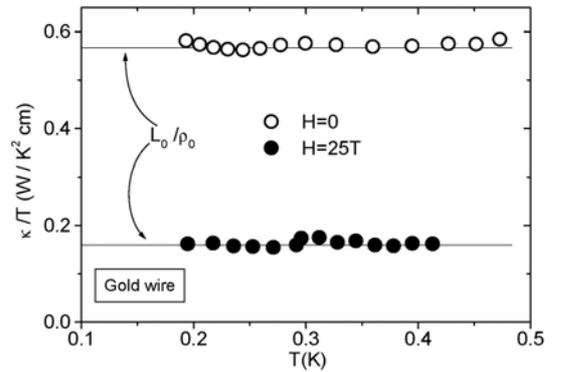

Fig.3 - Thermal conductivity of a gold wire at zero-field (open circles) and in presence of a 25T magnetic field (solid circles) compared to the expected magnitudes according to the Wiedemann-Franz law.

The cuprate superconductor $Bi_{2+x}Sr_{2-x}CuO_{6+\delta}$ (Bi-2201) was the first superconductor to be investigated by this set-up. Compared to other families of high-$T_c$ cuprates, the maximum critical temperature obtained in this compound is relatively low (~10K) and the upper critical field (~25T) lies within the range accessible by a resistive DC magnet [15]. Note that current commercial superconducting magnets cannot provide a field of such strength.

Fig.4 displays the temperature dependence of thermal conductivity in absence and in presence of a strong magnetic field. Here, a sizeable part of heat conductivity is due to the contribution of phonons. The electronic contribution can be estimated by extracting a finite κ/T in the T=0 limit [10]. Note that the lowest data point obtained at H=25T corresponds to T=0.145K.

As seen in the inset of the figure, a magnetic field of 25T is strong enough to destroy any trace of superconductivity in the electrical resistivity of the sample. Strikingly, however, the application of such a magnetic field leaves the low-temperature thermal conductivity virtually unchanged. Therefore, the magnitude of electronic heat conductivity is almost identical in the normal and in the superconducting states. This surprising result provides fresh input for the ongoing debate on heat transport by the nodal quasi-particles of the d-wave superconductor and on the nature of the elementary excitations of the normal state. A more detailed discussion of the result is presented in ref. 10.

In summary, we designed and realized a new set-up in order to measure thermal conductivity in sub-Kelvin temperature range in a resistive DC magnet. The two major technical challenges were to cool down the sample in spite of mechanical vibrations and to perform an accurate determination of temperature. The first series of experiments indicate that both problems were successfully resolved. Thermal conductivity measurements were carried out below 0.15K and a compelling precision on absolute magnitude of thermal conductivity was attained thanks to Coulomb Blockade Thermometry.

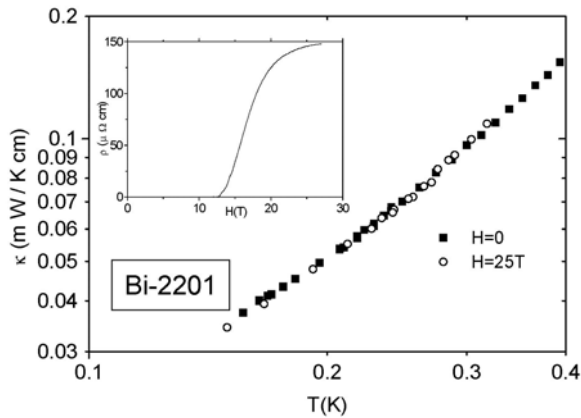

Fig. 4 – Thermal conductivity of a Bi-2201 sample at zero-magnetic field and at H=25T. The application of the magnetic filed leaves thermal conductivity virtually unchanged. Inset displays the resistive transition of the same sample at T=0.3K.

## Acknowledgements

The authors would like to thank Duncan Maude for his suggestion to do the measurement this way, his help with noise reduction and for reading the manuscript. We are also grateful to our other collaborators in the first experiment using this set-up: Sergey Vedeneev for providing the first sample to be measured; Cyril Proust for mounting the sample and Romain Bel for helping with data acquisition and analysis. Finally we acknowledge the precious assistance of Christian Gianese (CRTBT-CNRS) who made the silver sinters.